%% file: main.tex
\documentclass[letterpaper, 10pt, conference]{ieeeconf}
\IEEEoverridecommandlockouts
\usepackage{decar-common}
\usepackage{decar-post}

\makeatletter
\AtBeginDocument{
  \check@mathfonts
}
\makeatother

\usepackage{standalone}
\usepackage{tikz}
\usepackage{svg}
\usepackage{pgfplots}
\usetikzlibrary{positioning}
\usetikzlibrary{calc}

\title{\LARGE \bf \texttt{dkpy}: Robust Control with Structured Uncertainty in 
Python}

\author{Timothy Everett Adams, Steven Dahdah, and James Richard 
Forbes$^{1}$
\thanks{This work was supported by the NSERC Discovery Grant Program, FRQNT,
CIFAR, and the CRM.}
\thanks{$^{1}$The authors are with the  Department of Mechanical Engineering,
McGill University, 817 Sherbrooke St. W., Montreal, QC H3A 0C3, Canada.
The corresponding author is {\tt\small timothy.adams@mail.mcgill.ca}.}
}

\begin{document}

\maketitle

% Abstract
\begin{abstract}
    Models used for control design are, to some degree, uncertain.
    Model uncertainty must be accounted for to ensure the robustness of the
    closed-loop system.
    $\mu$-analysis and $\mu$-synthesis methods allow for the analysis and
    design of controllers subject to structured uncertainties.
    Moreover, these tools can be applied to robust performance problems as they
    are fundamentally robust control problems with structured uncertainty.
    The contribution of this paper is \texttt{dkpy}, an open-source Python
    package for performing robust controller analysis and synthesis for systems
    subject to structured uncertainty.
    \texttt{dkpy} also provides tools for performing model uncertainty
    characterization using data from a set of perturbed systems.
    The open-source project can be found at \\
    \centerline{\texttt{\url{https://github.com/decargroup/dkpy}}.}
\end{abstract}

% Keywords
\begin{keywords}
    Robust control, structured uncertainty,\\ $\mu$-analysis, $DK$-iteration,
    Python
\end{keywords}

\section{Introduction}
\label{sec:introduction}
There is always uncertainty in models used for control design due to unmodeled
dynamics, operating condition variations, and uncertain parameters.
In many applications, it is crucial to model this uncertainty and account for
it in the control design process to ensure that the closed-loop system is
stable and achieves the prescribed level of performance for all expected plant
variations.
A common model description used for control design is the linear time-invariant
(LTI) system \cites{antsaklisLinearSystemsPrimer2007,
greenLinearRobustControl2012, skogestadMultivariableFeedbackControl2005,
williamsLinearStateSpaceControl2007}.
LTI models often arise from the linearization of nonlinear systems or by
neglecting nonlinear effects in the modelling process
\cite{antsaklisLinearSystemsPrimer2007}.
Therefore, the uncertainty in LTI models must be accurately characterized in
order to realize control systems that are robust to the errors in these
simplified models.

Often the uncertainty in complex interconnected models is structured because
each subsystem generates a specific form of uncertainty.
The structure in the uncertainty can be leveraged to reduce the conservatism in
the robustness characterization that would otherwise be present if the
structure were neglected.
The main method used for the analysis of LTI systems with structured
uncertainty is the structured singular value (SSV), often denoted as $\mu$
\cite{packardComplexStructuredSingular1993},
\cite[\S~8.8]{skogestadMultivariableFeedbackControl2005},
\cite[\S~10.2]{zhouEssentialsRobustControl1998}.
Given its utility in robust control problems, it is crucial that there is
robust and user-friendly software that implements these $\mu$-analysis and
$\mu$-synthesis methods for control engineers to apply in their respective
domains.

MATLAB's \textit{Robust Control Toolbox} \cite{RobustControlToolbox} is a
mature software package that implements many robust control methods using the
SSV in a simple-to-use interface.
However, MATLAB is closed-source, limiting the availability of these powerful
tools to a wider range of users.
An open-source alternative for robust control tools would be valuable to the
control community as it would allow for the development and exchange of ideas
in a shared space.

The contribution of this paper is the development of \texttt{dkpy}, an
open-source Python package for performing robust control analysis and synthesis
with structured uncertainty.
\texttt{dkpy} is built on the \texttt{python-control} library
\cite{fullerPythonControlSystems2021} and \texttt{slycot}
\cite{PythoncontrolSlycot2025} in order to integrate with the rest of the
open-source Python control systems ecosystem.
At the time of writing, \texttt{dkpy} \texttt{v0.1.9} implements the following
features:
\begin{packed_itemize}
    \item Robust closed-loop analysis with complex structured uncertainty using
        $\mu$-analysis;
    \item Robust controller synthesis with complex structured uncertainty using
        $DK$-iteration;
    \item Multi-model unstructured uncertainty characterization.
\end{packed_itemize}

The remainder of the paper is structured as follows.
Section~\ref{sec:preliminaries} discusses the mathematical background for
robust control analysis and synthesis using the SSV.
Section~\ref{sec:implementation} discusses the software architecture of
\texttt{dkpy} as well as the currently implemented functionality of the
package.
Section~\ref{sec:examples} showcases the use of \texttt{dkpy} in standard
robust control examples.
Finally, Section~\ref{sec:conclusion} concludes the paper.

\section{Preliminaries}
\label{sec:preliminaries}

\subsection{Signals, Systems, and Norms}
\label{sec:signals_systems_norms}

$\mc{L}^{n}_{2} = \mc{L}^{n}_{2}[0, \infty]$ is the one-sided infinite horizon
Lebesgue 2-space of signals, where $n$ denotes the dimension of the signal
\cite[\S~3.1]{greenLinearRobustControl2012}.
$\mbc{G}: \mc{L}^{m}_{2} \rightarrow \mc{L}^{n}_{2}$ is an LTI system with a
transfer function representation $\mbf{G}(s): \cnums \rightarrow \cnums^{m
\times n}$ \cite[\S~3.2.1]{greenLinearRobustControl2012}.
The $\mc{H}_{\infty}$-norm of an LTI system in terms of its transfer function
is defined as \cite[\S~3.2.3]{greenLinearRobustControl2012}
\begin{align}
    \norm{\mbc{G}}_{\infty} = \sup_{\omega} \bar{\sigma}(\mbf{G}(j\omega)),
    \label{eq:hinf_norm}
\end{align}
and $\mc{H}_{\infty}$ denotes the class of systems that are analytic in the
open-right-half plane and have a bounded $\mc{H}_{\infty}$-norm.
Moreover, $\mc{RH}_{\infty}$ denotes the real rational subspace of
$\mc{H}_{\infty}$.

\subsection{Robust Stability with Structured Uncertainty}
\label{sec:robust_stability}

\begin{figure}[t]
    \vspace{5pt}
    \centering
    \begin{subfigure}{0.45\linewidth}
        \centering
        \input{figs/block_diagram/robust_stability_synthesis.tex}
        \caption{Robust control synthesis.}
        \label{fig:robust_control_synthesis_diagram}
    \end{subfigure}
    \hfill
    \begin{subfigure}{0.45\linewidth}
        \centering
        \input{figs/block_diagram/robust_stability_analysis.tex}
        \caption{Robust control analysis.}
        \label{fig:robust_control_analysis_diagram}
    \end{subfigure}
    \caption{General robust control diagrams.}
    \vspace{-15pt}
    \label{fig:robust_control_diagram}
\end{figure}
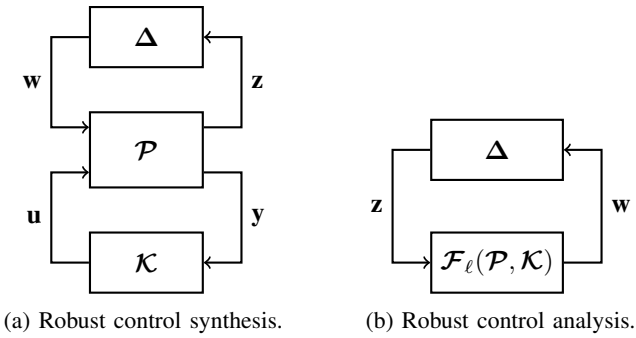

Figure~\ref{fig:robust_control_synthesis_diagram} shows the block diagram of
the standard robust stability problem.
In this formulation, a generalized plant $\mbc{P}$ relates the exogenous inputs
$\mbf{w} \in \mc{L}^{n_{w}}_{2}$ and controller outputs $\mbf{u} \in
\mc{L}^{n_{u}}_{2}$ to the exogenous outputs $\mbf{z} \in \mc{L}^{n_{z}}_{2}$
and controller inputs $\mbf{y} \in \mc{L}^{n_{y}}_{2}$.
In transfer function form, the relationship is
\begin{align}
    \begin{bmatrix}
        \mbf{z}(s) \\
        \mbf{y}(s)
    \end{bmatrix}
    =
    \underbrace{
    \begin{bmatrix}
        \mbf{P}_{11}(s) & \mbf{P}_{12}(s) \\
        \mbf{P}_{21}(s) & \mbf{P}_{22}(s)
    \end{bmatrix}
    }_{\mbf{P}(s)}
    \begin{bmatrix}
        \mbf{w}(s) \\
        \mbf{u}(s)
    \end{bmatrix}.
    \label{eq:generalized_plant}
\end{align}
The controller inputs $\mbf{y}$ are mapped to the controller outputs $\mbf{u}$
via the controller $\mbc{K}$, which in transfer function form is 
\begin{align}
    \mbf{u}(s) = \mbf{K}(s) \mbf{y}(s).
    \label{eq:controller}
\end{align}
The feedback interconnection of $\mbf{P}(s)$ and $\mbf{K}(s)$ can be written
using the lower linear fractional transformation (LFT).
The lower LFT between these systems is
\cite[\S~8.1]{skogestadMultivariableFeedbackControl2005}
\begin{align}
    &\mbf{F}_{\ell}(\mbf{P}, \mbf{K})(s) = \nonumber \\
    &\qquad
    \mbf{P}_{11}(s)
    + \mbf{P}_{12}(s) \mbf{K}(s) 
    (\eye - \mbf{P}_{22}(s) \mbf{K}(s))\inv \mbf{P}_{21}(s).
\end{align}
The uncertain feedback interconnection with the lower LFT is depicted in
Figure~\ref{fig:robust_control_analysis_diagram}.
The exogenous inputs $\mbf{w}$ and outputs $\mbf{z}$ are related through an LTI
structured perturbation $\mbs{\Delta}$ via
\begin{align}
    \mbf{w}(s) = \mbs{\Delta}(s) \mbf{z}(s).
    \label{eq:delta_block}
\end{align}
The uncertainty structure is defined by a complex block diagonal matrix set
\cite[\S~10.2]{zhouEssentialsRobustControl1998}
\begin{align}
    &\mbs{\Gamma}^{\omega} = \nonumber \\
    &\quad
    \left\{ 
        \diag \left(
        \{ \delta_{i} \eye_{r_{i}} \},\,
        \{ \mbs{\Delta}_{j} \}
        \right)
        \, \big| \,
        \delta_{i} \in \cnums,\,
        \mbs{\Delta}_{j} \in \cnums^{m_{j} \times m_{j}}
    \right\}.
    \label{eq:uncertainty_structure_set}
\end{align}
where $\delta_{i} \eye_{r_{i}}$ for $i \in \{ 1, \dots, M \}$ represent
diagonal repeated uncertainty blocks and $\mbs{\Delta}_{j}$ for $j \in \{ 1,
\dots, F \}$ represent full uncertainty blocks.
The set $\mbs{\Gamma}^{\omega}$ describes the uncertainty structure in the form
of complex block diagonal matrices whereas the perturbation $\mbs{\Delta}$ must
be an LTI system.
Therefore, the set of perturbation systems is defined as
\cite[\S~10.3]{zhouEssentialsRobustControl1998}
\begin{align}
    \mbs{\Gamma} =
    \left\{ \mbs{\Delta}(s) \in \mc{RH}_{\infty}
    \, \Bigg| \,
    \begin{aligned}
        & \norm{\mbs{\Delta}(s)}_{\infty} \leq 1, \\
        & \mbs{\Delta}(j\omega) \in \mbs{\Gamma}_{\omega} \; \forall \omega \in \rnums
    \end{aligned}
    \right\},
    \label{eq:pertubation_set}
\end{align}
which corresponds to real rational LTI perturbations with
$\mc{H}_{\infty}$-norm less than or equal to $1$ with block diagonal structure
described by $\mbs{\Gamma}^{\omega}$.

The robust stability problem is to assess whether the uncertain feedback
interconnection, described by \eqref{eq:generalized_plant},
\eqref{eq:controller}, and \eqref{eq:delta_block}, and shown in
Figure~\ref{fig:robust_control_diagram}, is internally stable for all
perturbations $\mbs{\Delta}$ in the uncertainty set $\mbs{\Gamma}$
\cite[\S~8.4]{skogestadMultivariableFeedbackControl2005}.

\subsection{Robust Performance}
\label{sec:robust_performance}

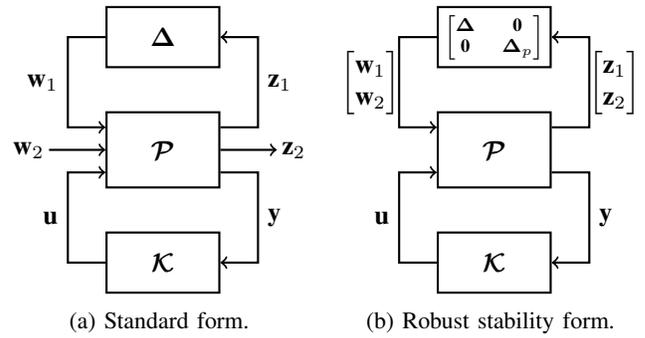
\begin{figure}[t]
    \vspace{5pt}
    \centering
    \begin{subfigure}[t]{0.49\linewidth}
        \centering
        \input{figs/block_diagram/robust_performance_1.tex}
        \caption{Standard form.}
        \label{fig:robust_performance_standard_form}
    \end{subfigure}
    \hfill
    \begin{subfigure}[t]{0.49\linewidth}
        \centering
        \input{figs/block_diagram/robust_performance_2.tex}
        \caption{Robust stability form.}
        \label{fig:robust_performance_stability_form}
    \end{subfigure}
    \caption{Equivalent forms of the robust performance problem diagrams.}
    \vspace{-15pt}
    \label{fig:robust_performance}
\end{figure}

The robust performance problem is to assess whether a closed-loop performance
specification is met for all possible uncertainty perturbations.
Figure \ref{fig:robust_performance_standard_form} shows the standard block
diagram for the robust performance problem.
There are a set of signals $\mbf{w}_{1} \in
\mc{L}_{2}^{n_{w_1}}, \mbf{z}_{1} \in \mc{L}_{2}^{n_{z_1}}$ in feedback with
the uncertainty $\mbs{\Delta}$ and a set of signals $\mbf{w}_{2} \in
\mc{L}_{2}^{n_{w_1}}, \mbf{z}_{2} \in \mc{L}_{2}^{n_{w_2}}$ related to the
closed-loop performance that are passed through the generalized
plant.
$\mbf{w}_{2}$ may include reference tracking signals, disturbances and sensor
noise whereas $\mbf{z}_{2}$ may include tracking errors and actuator effort.
The performance specification that must be met for all perturbations is given
by the closed-loop gain from $\mbf{w}_{2}$ to $\mbf{z}_{2}$ measured using the
$\mc{H}_{\infty}$-norm.
\cite[\S~8.4]{skogestadMultivariableFeedbackControl2005}.
Note that the robust performance problem is distinct from the optimal
performance problem subject to a robust stability constraint.
The former guarantees the performance across all uncertainties whereas the
latter only guarantees it for the nominal case.

The robust performance problem has an equivalent robust stability form.
Consider a fictitious unstructured perturbation $\mbs{\Delta}_{p}
\in \mc{RH}_{\infty}$ such that $\norm{\mbs{\Delta}_p}_{\infty} \leq 1$ and 
\begin{align}
    \mbf{w}_{2}(s) = \mbs{\Delta}_{p}(s) \mbf{z}_{2}(s),
    \label{eq:delta_block_performance}
\end{align}
which results in the feedback interconnection in
Figure~\ref{fig:robust_performance_stability_form}.
This is a special case of the robust stability problem with structured
uncertainty shown in Figure~\ref{fig:robust_control_synthesis_diagram}.
In fact, the robust stability of the feedback interconnection in
Figure~\ref{fig:robust_performance_stability_form} implies the robust
performance of that shown in Figure~\ref{fig:robust_performance_standard_form}
\cite[Theorem 10.8]{zhouEssentialsRobustControl1998}.
Therefore, the robust performance problem may be solved using the same tools 
as the structured uncertainty robust stability problem.

\subsection{Structured Singular Value and $\mu$-analysis}
\label{sec:mu_analysis}

The SSV, often denoted as $\mu$, is used in the robust stability analysis of
LTI systems with structured uncertainty.
The SSV yields less conservative results compared to analysis with unstructured
uncertainty \cite{skogestadMultivariableFeedbackControl2005}.
The SSV can not be computed directly.
Rather, upper and lower bounds are used in its place.
The SSV upper bound for $\mbf{M} \in \cnums^{m \times n}$ and $\mbs{\Delta} \in
\mbs{\Gamma}^{\omega}$ is given by
\cite[\S~10.2.2]{zhouEssentialsRobustControl1998}
\begin{align}
    \mu_{\mbs{\Delta}}(\mbf{M}) \leq
    \min_{\mbf{D}_{\omega} \in \mbc{D}^{\omega}_{\mbs{\Delta}}}
    \bar{\sigma}\left(\mbf{D}_{\omega} \mbf{M} \mbf{D}_{\omega}\inv\right),
    \label{eq:ssv_upper_bound}
\end{align}
where $\mu_{\mbs{\Delta}}(\mbf{M})$ is the SSV of $\mbf{M}$ with respect to
$\mbs{\Delta} \in \mbs{\Gamma}^{\omega}$, and $\mbc{D}^{\omega}_{\mbs{\Delta}}$
is the set of matrices that commute with $\mbs{\Delta}$ such that
\begin{align}
    \mbc{D}_{\mbs{\Delta}}^{\omega} = 
    \left\{
        \mbf{D} \in \cnums^{n \times n}
         \, \Big| \, 
         \mbf{D} \mbs{\Delta} = \mbs{\Delta} \mbf{D},\,
         \mbs{\Delta} \in \mbs{\Gamma}^{\omega}
    \right\}.
    \label{eq:d_scale_matrix_set}
\end{align}

The SSV is used to assert the robust stability of an uncertain feedback
interconnection.
The interconnection shown in Figure~\ref{fig:robust_control_analysis_diagram}
is well-posed and internally stable for all $\mbs{\Delta}(s) \in \mbs{\Gamma}$
if and only if \cite[Theorem 10.7]{zhouEssentialsRobustControl1998}
\begin{align}
    \sup_{\omega \in \rnums}
    \mu_{\mbs{\Delta}}(\mbf{F}_{\ell}(\mbf{P}, \mbf{K})(j\omega)) \leq 1.
    \label{eq:robust_stability}
\end{align}
In practice, the upper bound in \eqref{eq:ssv_upper_bound} is used and the
condition is evaluated over a grid of frequencies $\Omega = \{\omega_{1},
\dots, \omega_{N}\}$.
Therefore, the robust stability condition used in practice is 
\begin{align}
    \min_{\mbf{D}_{\omega} \in \mbc{D}^{\omega}_{\Delta}}
    \bar{\sigma}
    \left(
        \mbf{D}_{\omega}
        \mbf{F}_{\ell}(\mbf{P}, \mbf{K})(j\omega)
        \mbf{D}_{\omega}\inv
    \right)
    \leq 1 \; \forall \omega \in \Omega.
    \label{eq:robust_stability_practical}
\end{align}

\subsection{$DK$-iteration and $\mu$-synthesis}
\label{sec:mu_synthesis}

A robust controller $\mbc{K}$ can be synthesized by minimizing the upper bound
of the SSV.
The robust stability condition in \eqref{eq:robust_stability_practical} is
defined over a frequency grid whereas standard controller synthesis methods
require continuous LTI systems.
Therefore, the scaling matrix response $\mbf{D}_{\omega}$ must be converted to
an LTI system $\mbf{D}(s)$ and the singular value over the grid of frequencies
is replaced with the $\mc{H}_{\infty}$-norm.
This leads to the optimal control design problem
\cite[\S~10.4]{zhouEssentialsRobustControl1998}
\begin{align}
    \min_{\mbf{K}(s)}
    \left(
        \min_{\mbf{D}(s) \in \mbc{D}_{\mbs{\Delta}}}
        \norm{
            \mbf{D}(s) \mbf{F}_{\ell}(\mbf{P}, \mbf{K})(s) \mbf{D}\inv(s) 
        }_{\infty}
    \right),
    \label{eq:mu_synthesis}
\end{align}
where $\mbc{D}_{\mbs{\Delta}}$ is the set of stable rational LTI systems with
stable rational inverses that commute with the perturbation $\mbs{\Delta}(s)$
defined by
\begin{align}
    \mbc{D}_{\mbs{\Delta}} =
    \left\{
    \mbf{D}(s) \in \mc{RH}_{\infty}
    \,\Bigg|\, 
    \begin{aligned}
    &\mbf{D}\inv(s) \in \mc{RH}_{\infty} \\
    &\mbf{D}(j\omega) \in \mbc{D}^{\omega}_{\mbs{\Delta}} \;
        \forall \omega \in \rnums
    \end{aligned}
    \right\}.
    \label{eq:d_scale_set}
\end{align}
The optimization problem in \eqref{eq:mu_synthesis} is non-convex in the
variables $\mbf{K}(s)$ and $\mbf{D}(s)$.
However, \eqref{eq:mu_synthesis} is convex in each of the design variables if
the other is held fixed.
This leads to a heuristic controller synthesis method known as $DK$-iteration
in which the objective in \eqref{eq:mu_synthesis} is minimized by alternating
between $\mbf{K}(s)$ and $\mbf{D}(s)$ while holding the other fixed.
The $DK$-iteration procedure is as follows:
\begin{packed_enum}\setcounter{enumi}{-1}
    \item\label{item:dk_iteration_init} Initialize the $\mbf{D}$-scale with any
        $\mbf{D}(s) \in \mbc{D}_{\mbs{\Delta}}$, for example $\mbf{D}(s) =
        \mbf{1}$.
    \item\label{item:dk_iteration_hinf} Synthesize a controller $\mbf{K}(s)$
        for the scaled closed-loop system with fixed scales $\mbf{D}(s)$
        \begin{align}
            \min_{\mbf{K}(s)}
                \norm{
                    \mbf{D}(s)
                    \mbf{F}_{\ell}(\mbf{P}, \mbf{K})(s)
                    \mbf{D}\inv(s) 
                }_{\infty}.
                \label{eq:dk_iteration_hinf}
        \end{align}
    \item\label{item:dk_iteration_ssv} Compute the upper bound on the SSV
        $\bar{\mu}_{\mbs{\Delta}}(j\omega)$ and scaling matrix frequency
        response $\mbf{D}(j\omega)$ over the discrete grid of frequencies
        $\Omega = \{\omega_{1}, \dots, \omega_{N}\}$
        \begin{align}
            \bar{\mu}_{\mbs{\Delta}}(j\omega) =
            \min_{\mbf{D}_{\omega} \in \mbc{D}^{\omega}_{\Delta}}
            \bar{\sigma}
            \left(
                \mbf{D}_{\omega}
                \mbf{F}_{\ell}(\mbf{P}, \mbf{K})(j\omega)
                \mbf{D}_{\omega}\inv
            \right).
        \end{align}
        If the robust stability condition in
        \eqref{eq:robust_stability_practical} is met, stop. Otherwise,
        continue.
    \item\label{item:dk_iteration_fit} Fit a stable and minimum-phase transfer
        function $\mbf{D}(s)$ of a given order to the magnitude response of
        $\mbf{D}_{\omega}$ for $\omega \in \Omega$. Return to step
        \ref{item:dk_iteration_hinf}).
\end{packed_enum}

\subsection{Uncertainty Characterization}
\label{sec:uncertainty_characterization}

In many applications, a set of perturbed models is used to describe
the uncertainty associated with a nominal LTI model $\mbf{G}_{0}(s)$.
In many situations, unstrutured uncertainty models are sufficient to describe
the off-nominal models.
For example, the multiplicative input uncertainty model defines a set of
perturbed systems as
\begin{align}
    \mbc{G}_{I} =
    \left\{
        \mbf{G}_{0}(s)(\eye + \mbf{E}_{I}(s))
        \;\Bigg|\;
        \begin{aligned}
            &\mbf{E}_{I}(s) = \mbf{W}_{L}(s) \mbs{\Delta}(s) \mbf{W}_{R}(s), \\
            &||\mbs{\Delta}||_{\infty} \leq 1
        \end{aligned}
    \right\},
    \label{eq:multiplicative_input_uncertainty_set}
\end{align}
where $\mbf{E}_{I}(s)$ is the residual, $\mbs{\Delta}(s) \in \mc{RH}_{\infty}$
is a normalized unstructured perturbation, and $\mbf{W}_{L}(s), \mbf{W}_{R}(s)
\in \mc{RH}_{\infty}$ are uncertainty weights
\cite[\S~8.3.4]{zhouEssentialsRobustControl1998}.
Additional unstructured uncertainty models can be found in
\cite[\S~8.2.3]{skogestadMultivariableFeedbackControl2005},
\cite[\S~8.3.4]{zhouEssentialsRobustControl1998}.

One method to construct an uncertainty set of the form
\eqref{eq:multiplicative_input_uncertainty_set} is to use frequency response
data of a nominal model $\mbf{G}_{0}(j\omega)$ and a series of perturbed
off-nominal models $\mbf{G}_{k}(j\omega)$ for $k \in \{1, \dots, K\}$ and
$\omega \in \{\omega_{1}, \dots, \omega_{N}\}$.
This process is known as multi-model uncertainty characterization.
The steps required to construct the uncertainty set are:
\begin{packed_enum}
\item Compute the residual response $\mbf{E}_{(\cdot), k}(j\omega)$ from 
    $\mbf{G}_{0}(j\omega)$ and $\mbf{G}_{k}(j\omega)$ for $k \in \{1, \dots,
    K\}$ and $\omega \in \{\omega_{1}, \dots, \omega_{N}\}$. For multiplicative
    input uncertainty, this is performed by solving the linear equations
        \begin{align}
            \mbf{G}_{0}(j\omega) \mbf{E}_{I, k}(j\omega) =
            (\mbf{G}_{k}(j\omega) - \mbf{G}_{0}(j\omega)).
            \label{eq:residual_multiplicative_input_uncertainty}
        \end{align}
    \item Compute the optimal uncertainty weight response
        $\mbf{W}_{L}(j\omega), \mbf{W}_{R}(j\omega)$ from
        $\mbf{E}_{(\cdot)}(j\omega)$ \cite{balasUncertainModelSet2009}.
    \item Fit overbounding, stable and minimum-phase LTI uncertainty weights
        $\mbf{W}_{L}(s), \mbf{W}_{R}(s)$ to the frequency responses
        $\mbf{W}_{L}(j\omega), \mbf{W}_{R}(j\omega)$.
\end{packed_enum}

\section{Implementation}
\label{sec:implementation}

\subsection{Robust Controller Analysis and Synthesis}
\label{sec:implementation_mu_analysis_synthesis}

\begin{figure}[t]
    \vspace{5pt}
    \centering
    \includegraphics[width=0.9\linewidth]{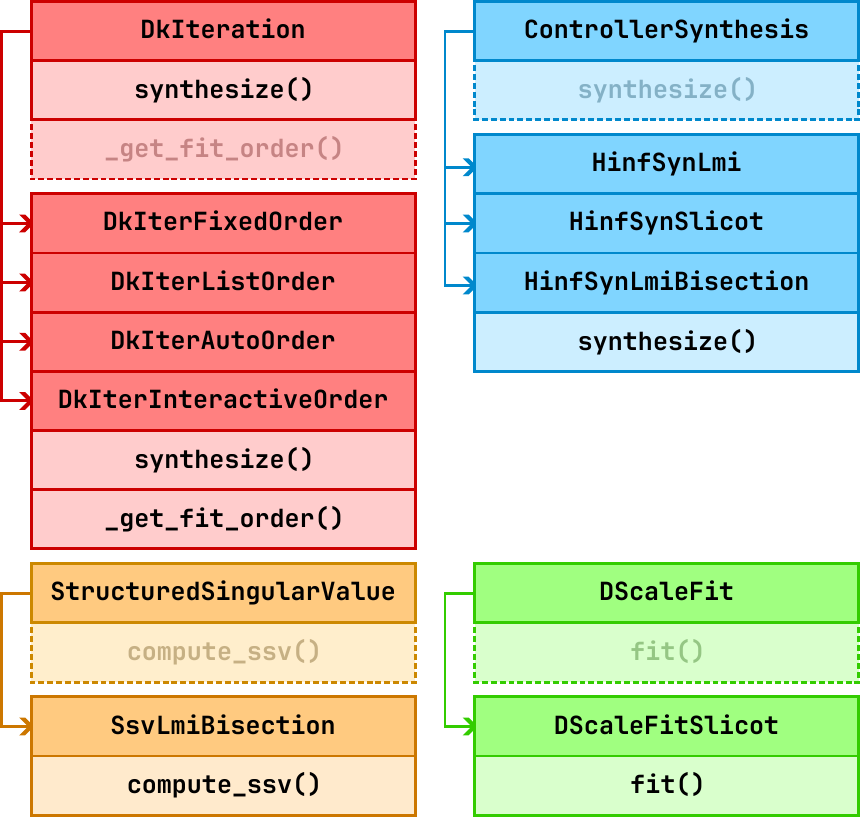}
    \caption{Inheritance structure of the abstract base classes in
    \texttt{dkpy}. Dark boxes denote a class, light boxes with a solid outline
    denote implemented methods, and light boxes with dashed outlines denote
    abstract (unimplemented) methods.}
    \vspace{-15pt}
    \label{fig:dkpy_abc_overview}
\end{figure}

\texttt{dkpy} implements three abstract base classes (ABCs) for the fundamental
processes required for $\mu$-analysis and $\mu$-synthesis.
ABCs provide implemented methods that are available to subclasses and
abstract methods that define required functionality that must be 
implemented by subclasses.
Therefore, an ABC provides a generic interface for some functionality in the
program.
Then, subclasses implement this functionality using their own approaches.
These implementations can be used interchangeably allowing for greater
modularity as well as the opportunity for users to define their own
implementations within the framework.
The fundamental ABCs and their implementations at the time of writing, which
are illustrated in Figure~\ref{fig:dkpy_abc_overview}, are
\begin{packed_enum}
    \item \texttt{ControllerSynthesis}: Synthesize a controller given the
        generalized plant with the \texttt{synthesize} abstract method,
        \begin{packed_itemize}
            \item \texttt{HinfSynSlicot}: $\mc{H}_{\infty}$-synthesis using the
                SLICOT routine \texttt{SB10AD} from \texttt{python-control}
                \cite{bennerSLICOTSubroutineLibrary1999,
                fullerPythonControlSystems2021},
            \item \texttt{HinfSynLmi}: $\mc{H}_{\infty}$-synthesis using an LMI
                formulation
                \cite[\S~V-A]{schererMultiobjectiveOutputfeedbackControl1997},
            \item \texttt{HinfSynLmiBisection}: $\mc{H}_{\infty}$-synthesis
                using a modified LMI formulation with bisection on the
                objective function.
                \cite[\S~V-A]{schererMultiobjectiveOutputfeedbackControl1997},
        \end{packed_itemize}
    \item \texttt{StructuredSingularValue}: Compute the SSV upper bound and
        associated scaling matrix frequency response given the closed-loop
        system frequency response and the perturbation block structure with the
        \texttt{compute\_ssv} abstract method,
        \begin{packed_itemize}
            \item \texttt{SsvLmiBisection}: LMI formulation with bisection on
                the SSV upper bound
                \cite[\S~4.25]{caverlyLMIPropertiesApplications2024},
        \end{packed_itemize}
    \item \texttt{DScaleFit}: Fit a stable and minimum-phase LTI system to the
        frequency response of the scaling matrices with the \texttt{fit}
        abstract method,
        \begin{packed_itemize}
            \item \texttt{DScaleFitSlicot}: Implementation using the SLICOT
                routine \texttt{SB10YD} from \texttt{slycot}
                \cite{bennerSLICOTSubroutineLibrary1999,
                PythoncontrolSlycot2025}.
        \end{packed_itemize}
\end{packed_enum}

The $\mu$-analysis process, outlined in Section~\ref{sec:mu_analysis}, is
performed using the \texttt{StructuredSingularValue} class.

The $\mu$-synthesis process using $DK$-iteration, described in
Section~\ref{sec:mu_synthesis}, is performed using an ABC called
\texttt{DkIteration}.
The \texttt{DkIteration} ABC is constructed from \texttt{ControllerSynthesis},
\texttt{StructuredSingularValue}, and \texttt{DScaleFit} implementations.
The \texttt{DkIteration} class implements a default method \texttt{synthesize}
that is inherited by all implementations.
The \texttt{synthesize} method takes a generalized plant and uncertainty
block structure as input and returns a robust controller along with information
about the solution, such as the SSV upper bound.
The \texttt{synthesize} method performs the iterative $DK$-iteration loop
described in Section~\ref{sec:mu_synthesis}.

\begin{figure}[t]
    \vspace{5pt}
    \centering
    \includegraphics[width=0.6\linewidth]{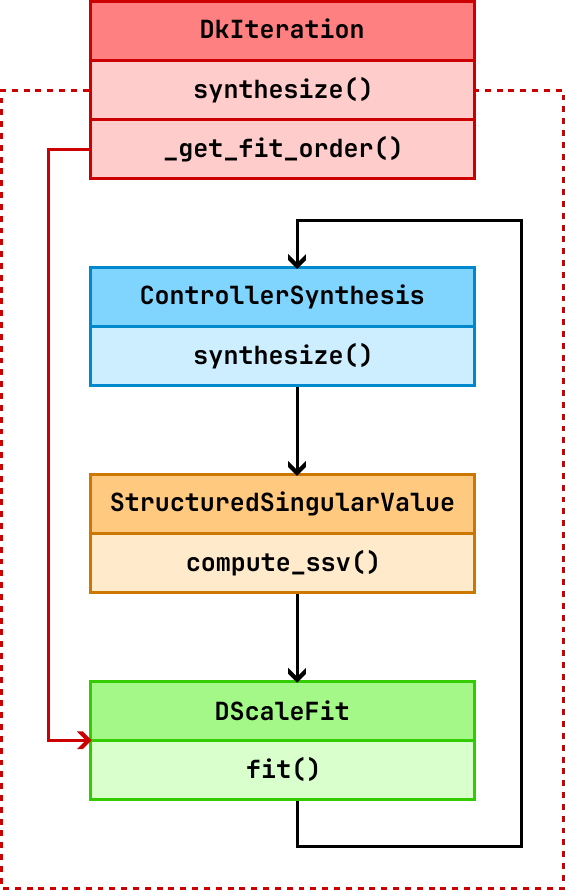}
    \caption{\texttt{DkIteration} \texttt{synthesize} method control flow for
    performing $DK$-iteration in \texttt{dkpy}.}
    \vspace{-15pt}
    \label{fig:dkpy_code_overview}
\end{figure}

The \texttt{DkIteration} ABC also defines an abstract method
\texttt{\_get\_fit\_order}, which determines the number of iterations and the
order of the scaling matrix fit at each iteration used by \texttt{DScaleFit}.
The \texttt{\_get\_fit\_order} implementation distinguishes the implementations
of \texttt{DkIteration} from one another.
Figure~\ref{fig:dkpy_code_overview} shows an overview of the controller
synthesis procedure using \texttt{DkIteration}.
The implementations of the \texttt{DkIteration} ABC at the time of writing,
which are also shown in Figure~\ref{fig:dkpy_abc_overview}, are
\begin{packed_enum}
    \item \texttt{DkIterFixedOrder}: Fixed number of iterations and fit order,
    \item \texttt{DkIterListOrder}:  Sequence of fit orders,
    \item \texttt{DkIterAutoOrder}: Automatically selected fit orders using the
        smallest error in the fitted frequency response,
    \item \texttt{DkIterInteractiveOrder}: Interactively selected fit orders
        from user prompts.
\end{packed_enum}

\subsection{Uncertainty Characterization}
\label{sec:implementation_uncertainty_characterization}

\begin{figure}[t]
    \vspace{5pt}
    \centering
    \includegraphics[width=0.7\linewidth]{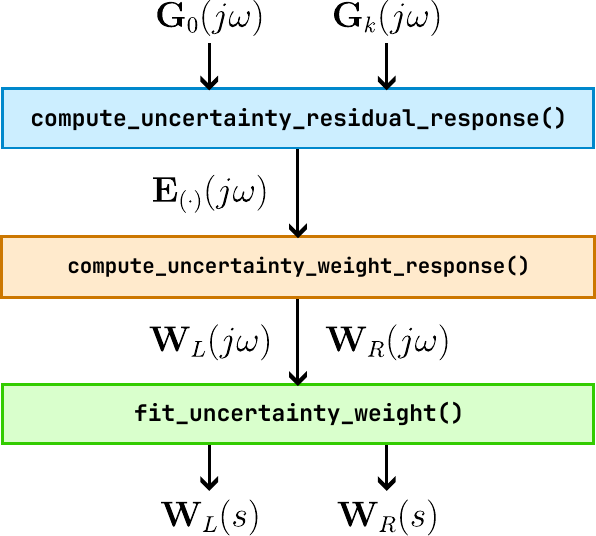}
    \caption{Multi-model uncertainty characterization code diagram in
    \texttt{dkpy}.}
    \vspace{-15pt}
    \label{fig:uncertainty_code_overview}
\end{figure}

\texttt{dkpy} provides utilities to perform multi-model uncertainty
characterization.
The code is organized around the three-step uncertainty characterization
process outlined in Section~\ref{sec:uncertainty_characterization}.
\texttt{dkpy} implements three functions, which are illustrated in
Figure~\ref{fig:uncertainty_code_overview}, to perform the uncertainty
characterization.
The functions are 
\begin{packed_enum}
    \item \texttt{compute\_uncertainty\_residual\_response}: Compute the
        residual frequency response through the linear equations of the form
        \eqref{eq:residual_multiplicative_input_uncertainty},
    \item \texttt{compute\_uncertainty\_weight\_response}: Compute the optimal
        weight frequency response subject to constraints on the weight
        structure using an LMI formulation \cite{balasUncertainModelSet2009}.
    \item \texttt{fit\_uncertainty\_weight}: Fit an overbounding, stable, and
        minimum-phase uncertainty weight to frequency response data using a
        log-Chebyshev method \cite[\S~5]{wuFIRFilterDesign1999}.
\end{packed_enum}

\section{Examples}
\label{sec:examples}

\begin{figure*}[t]
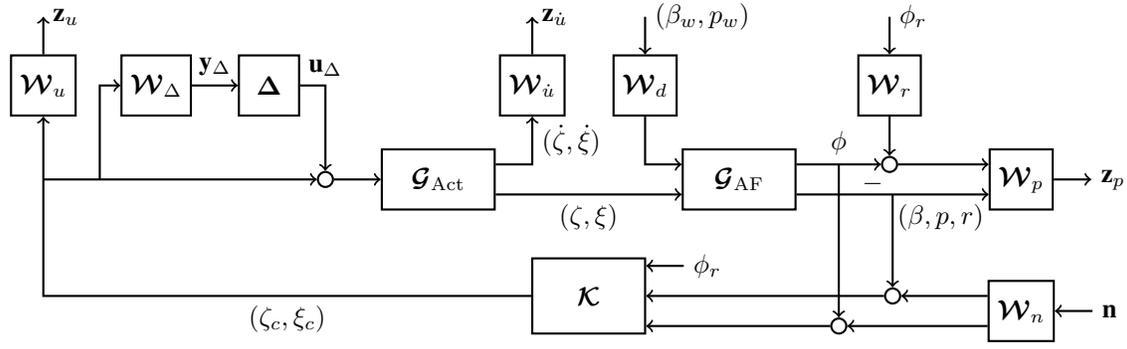

    \centering
    \includestandalone[scale=1.0]{figs/block_diagram/generalized_plant}
    \caption{Airframe lateral dynamics control problem block diagram.}
    \label{fig:generalized_plant_airframe}
    \vspace{-15pt}
\end{figure*}

The following examples are based on the aircraft control case study in
\cite[\S~14.1]{mackenrothRobustControlSystems2004}.
The lateral aircraft dynamics model in this example can be found in
\cite[\S~7.3, \S~B.4]{schmidtIntroductionAircraftFlight1998}.
Figure~\ref{fig:generalized_plant_airframe} shows the block diagram of the 
robust control problem and Table~\ref{tab:lateral_aircraft_parameters} provides
definitions of the signals.
The control objective is to track roll angle references with minimal sideslip
angle, roll rate, yaw rate, and actuator effort subject to wind disturbance and
sensor noise.

The control problem involves a model of the lateral aircraft dynamics
$\mbf{G}_{\mathrm{AF}}(s)$, a model of the rudder and aileron actuators
$\mbf{G}_{\mathrm{Act}}(s)$, a controller $\mbc{K}$, and various weights
$\mbf{W}_{(\cdot)}(s)$ that encode the bandwidths over which the disturbances
are expected and the performance variables should be minimized.
The variation arises from uncertainty in the actuator dynamics,
which is modeled as multiplicative input uncertainty.

% NOTE: Should the generalized plant model and weights be shown in a figure?

\subsection{Multi-model Uncertainty Characterization}
\label{sec:example_uncertainty_characterization}

\begin{table}[h]
    \vspace{5pt}
    \centering
    \caption{Lateral aircraft model variables.}
    \begin{tabular}{c c c}
        \hline \hline
        $\phi$ & Roll angle & $\si{deg}$ \\
        $\beta$ & Sideslip angle & $\si{deg}$ \\
        $p$ & Roll rate & $\si{deg/s}$ \\
        $r$ & Yaw rate & $\si{deg/s}$ \\\hline
        $\zeta$ & Rudder angle & $\si{deg}$ \\
        $\xi$ & Aileron angle & $\si{deg}$ \\
        $\zeta_{c}$ & Rudder angle command & $\si{deg}$ \\
        $\xi_{c}$ & Aileron angle command & $\si{deg}$ \\ \hline
        $\beta_{w}$ & Sideslip angle disturbance & $\si{deg}$ \\
        $p_{w}$ & Roll rate disturbance & $\si{deg/s}$ \\
        $n_{\phi}$ & Roll angle noise & $\si{deg}$ \\
        $n_{\beta}$ & Sideslip angle noise & $\si{deg}$ \\
        $n_{p}$ & Roll rate noise & $\si{deg/s}$ \\
        $n_{r}$ & Yaw rate noise & $\si{deg/s}$ \\\hline
        $\phi_{r}$ & Roll angle reference & $\si{deg}$ \\
        \hline \hline
    \end{tabular}
    \label{tab:lateral_aircraft_parameters}
    \vspace{-15pt}
\end{table}

In the case study, an uncertainty model and weight for the actuator dynamics
are given.
In this example, a new uncertainty model will be generated from a set of
perturbed actuator frequency responses to illustrate the uncertainty
characterization tools in \texttt{dkpy}.
The off-nominal actuators will be generated by sampling different perturbations
from the case study uncertainty model.

\begin{figure}[t]
    \centering
    \includegraphics[width=\columnwidth]{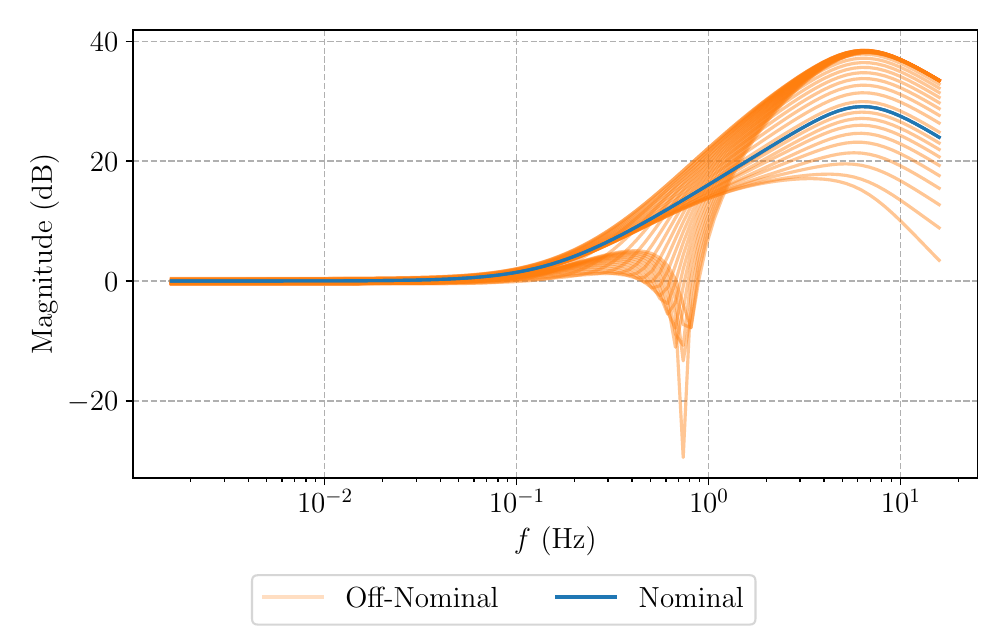}
    \caption{Singular value response of the nominal and off-nominal actuator
    models $\mbc{G}_{\mathrm{Act}}$.}
    \vspace{-15pt}
    \label{fig:uncertainty_example_sval_nom_offnom}
\end{figure}

A set of $40$ off-nominal actuators is generated using the multiplicative
input uncertainty model from the case study with different perturbations
$\norm{\mbs{\Delta}}_{\infty} \leq 1$.
Specifically, constant gain systems of the form
\begin{align*}
    \mbs{\Delta}(s) = a,\; a \in [-1, 1], 
\end{align*}
and all-pass filters of the form
\begin{align*}
    \mbs{\Delta}(s) = \f{\f{s}{b} - 1}{\f{s}{b} + 1},\; b \in [0.01, 10]
\end{align*}
are used.
Figure~\ref{fig:uncertainty_example_sval_nom_offnom} shows the singular value
response of the nominal and perturbed off-nominal actuator models.
The deviation between the actuator models increases with the frequency
indicating the presence of unmodeled high-frequency dynamics.

\begin{figure}[t!]
    \vspace{5pt}
    \centering
    \includegraphics[width=\columnwidth]{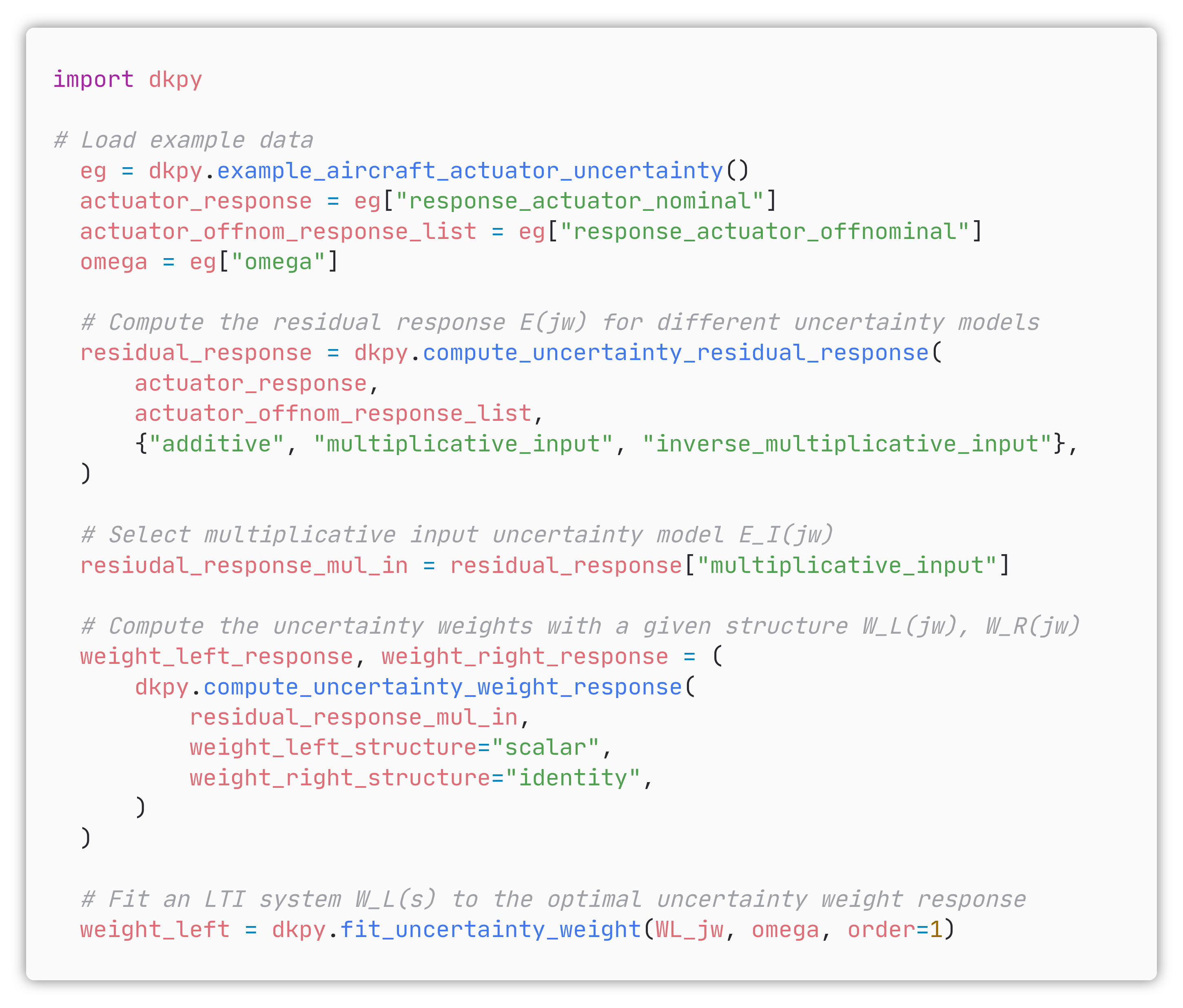}
    \caption{Uncertainty characterization code for an aircraft actuator model
    using \texttt{dkpy}.}
    \vspace{-15pt}
    \label{fig:uncertainty_example_code}
\end{figure}

\begin{figure}[t!]
    \centering
    \includegraphics[width=\columnwidth]{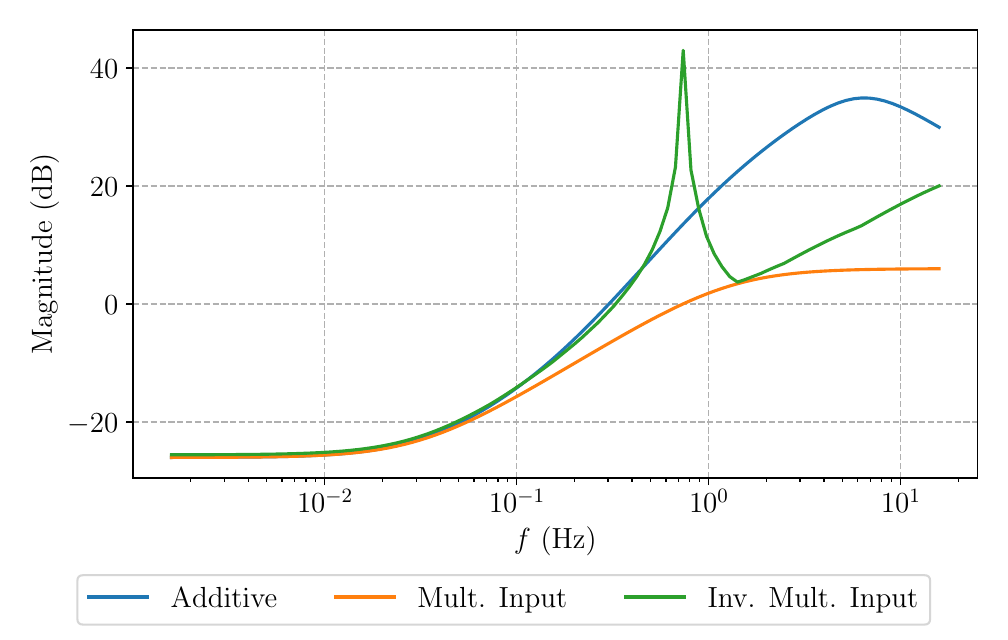}
    \caption{Singular value response of the uncertainty residuals for the
    uncertain actuator model for additive, multiplicative input, and inverse
    multiplicative input uncertainty.}
    \vspace{-15pt}
    \label{fig:uncertainty_example_residual_response}
\end{figure}

The uncertainty model is obtained using \texttt{dkpy} through the code shown in
Figure~\ref{fig:uncertainty_example_code}.
The residual response for additive, multiplicative input, and inverse
multiplicative input uncertainty models are computed from the nominal and
off-nominal actuator frequency response data.
The singular value response of the residuals for the different uncertainty
models is shown in Figure~\ref{fig:uncertainty_example_residual_response}.
The multiplicative input uncertainty model is selected as it yields the
smallest residual magnitude of the three alternatives.
This is no surprise given that the perturbed actuators were generated using a
multiplicative input uncertainty model to begin with.

\begin{figure}[t!]
    \centering
    \includegraphics[width=\columnwidth]{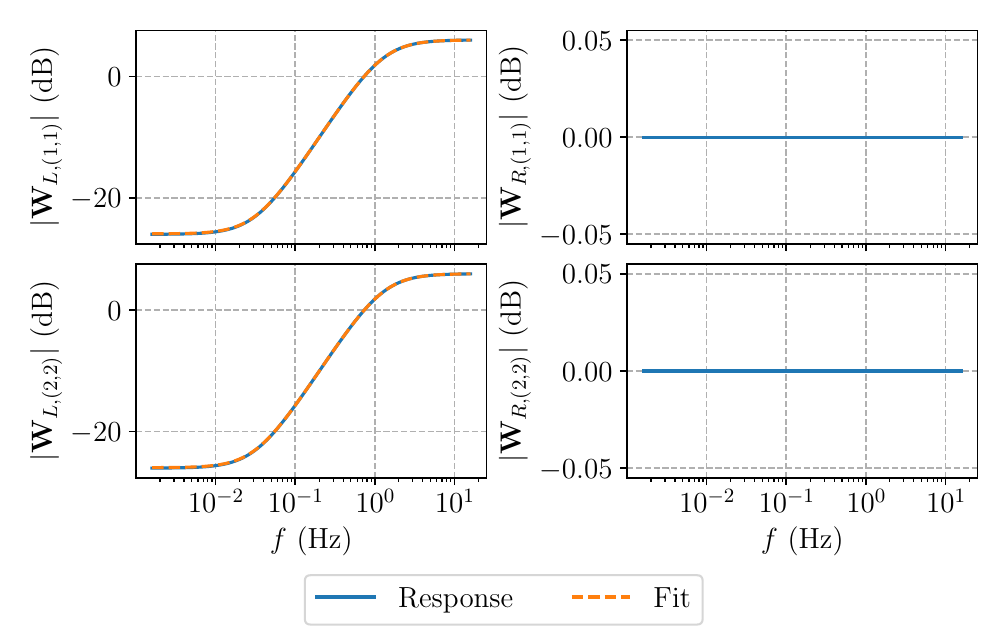}
    \caption{Magnitude response of the diagonal optimal uncertainty weight and
    the fitted overbounding stable and minimum-phase system. The weight
    $\mbf{W}_{R}(j\omega) = \mbf{1}$ is not fitted as it can be neglected in the
    generalized plant. }
    \vspace{-15pt}
    \label{fig:uncertainty_example_weight_response}
\end{figure}

The frequency response of the optimal uncertainty weights
$\mbf{W}_{L}(j\omega)$ and $\mbf{W}_{R}(j\omega)$ are computed.
The left uncertainty weight is constrained as $\mbf{W}_{L}(j\omega) =
w_{L}(j\omega) \eye$ and the right weight as
$\mbf{W}_{R}(j\omega) = \eye$ to mimic the uncertainty model used to generate
the off-nominal models.
Figure~\ref{fig:uncertainty_example_weight_response} shows the magnitude
response of the diagonal elements of the uncertainty weights.

Finally, an overbounding stable and minimum-phase LTI system $\mbf{W}_{L}(s)$
is fit to the magnitude data of the left uncertainty weight.
The right uncertainty weight fit is not performed as $\mbf{W}_{R}(j\omega) =
\mbf{1}$.
This means that $\mbf{W}_{R}(s)$ will not have any impact on the uncertainty
characterization nor the generalized plant.
Therefore, it can be neglected going forward.
Figure~\ref{fig:uncertainty_example_weight_response} also shows the magnitude
response of the fitted system in addition to the optimal frequency response.
It can be seen that the fitted weight tightly overbounds the optimal frequency
response indicating that there is little conservatism in the uncertainty
characterization.
%
% In fact, the uncertainty model obtained in this example using perturbed
% off-nominal actuators recovers the true uncertainty model used to generate the
% models.

\subsection{Robust Controller Synthesis}
\label{sec:example_robust_synthesis}

In this example, a controller that attains the robust performance criterion for
the control problem shown in Figure~\ref{fig:generalized_plant_airframe} is
designed.
As discussed in Section~\ref{sec:robust_performance}, the robust performance
problem is converted into a robust stability problem as in
Figure~\ref{fig:robust_performance_stability_form}.
In this problem, generalized plant inputs and outputs are
{\allowdisplaybreaks
\begin{align}
    \mbf{u}(s) &=
    \begin{bmatrix}
        \zeta_{c}(s) & \xi_{c}(s)
    \end{bmatrix}^{\trans}, \\
    \mbf{y}(s) &=
    \begin{bmatrix}
        \phi_{r}(s) & \phi(s) & \beta(s) & p(s) & r(s)
    \end{bmatrix}^{\trans}, \\
    \mbf{w}_{1}(s) &=
    \mbf{u}_{\Delta}(s) \\
    \mbf{z}_{1}(s) &=
    \mbf{y}_{\Delta}(s) \\
    \mbf{w}_{2}(s) &=
    \begin{bmatrix}
        \phi_{r}(s) & \mbf{n}^{\trans}(s) & \mbf{d}_{w}^{\trans}(s)
    \end{bmatrix}^{\trans}, \\
    \mbf{z}_{2}(s) &=
    \begin{bmatrix}
        \mbf{z}_{p}^{\trans}(s) & \mbf{z}_{u}^{\trans}(s) & \mbf{z}_{\dot{u}}^{\trans}(s)
    \end{bmatrix}^{\trans},
\end{align}}
where the noise is $\mbf{n}(s) = \bbm n_{\phi}(s) & n_{\beta}(s) & n_{p}(s) &
n_{r}(s) \ebm^{\trans}$ and the wind disturbance is $\mbf{d}_{w}(s) = \bbm
\beta_{w}(s) & p_{w}(s) \ebm^{\trans}$. 
The uncertain perturbation is $\mbshat{\Delta}$
\begin{align}
    \underbrace{\begin{bmatrix}
        \mbf{w}_{1}(s) \\
        \mbf{w}_{2}(s)
    \end{bmatrix}}_{\mbf{w}(s)}
    =
    \underbrace{\begin{bmatrix}
        \mbs{\Delta}(s) & \mbf{0} \\
        \mbf{0} & \mbs{\Delta}_{p}(s)
    \end{bmatrix}}_{\mbshat{\Delta}(s)}
    \underbrace{\begin{bmatrix}
        \mbf{z}_{1}(s) \\
        \mbf{z}_{2}(s)
    \end{bmatrix}}_{\mbf{z}(s)},
\end{align}
where $\mbs{\Delta}$ and $\mbs{\Delta}_{p}$ are full complex perturbations
associated with the multiplicative input uncertainty of the actuator and the
performance condition, respectively.

\begin{figure}[t]
    \centering
    \includegraphics[width=\columnwidth]{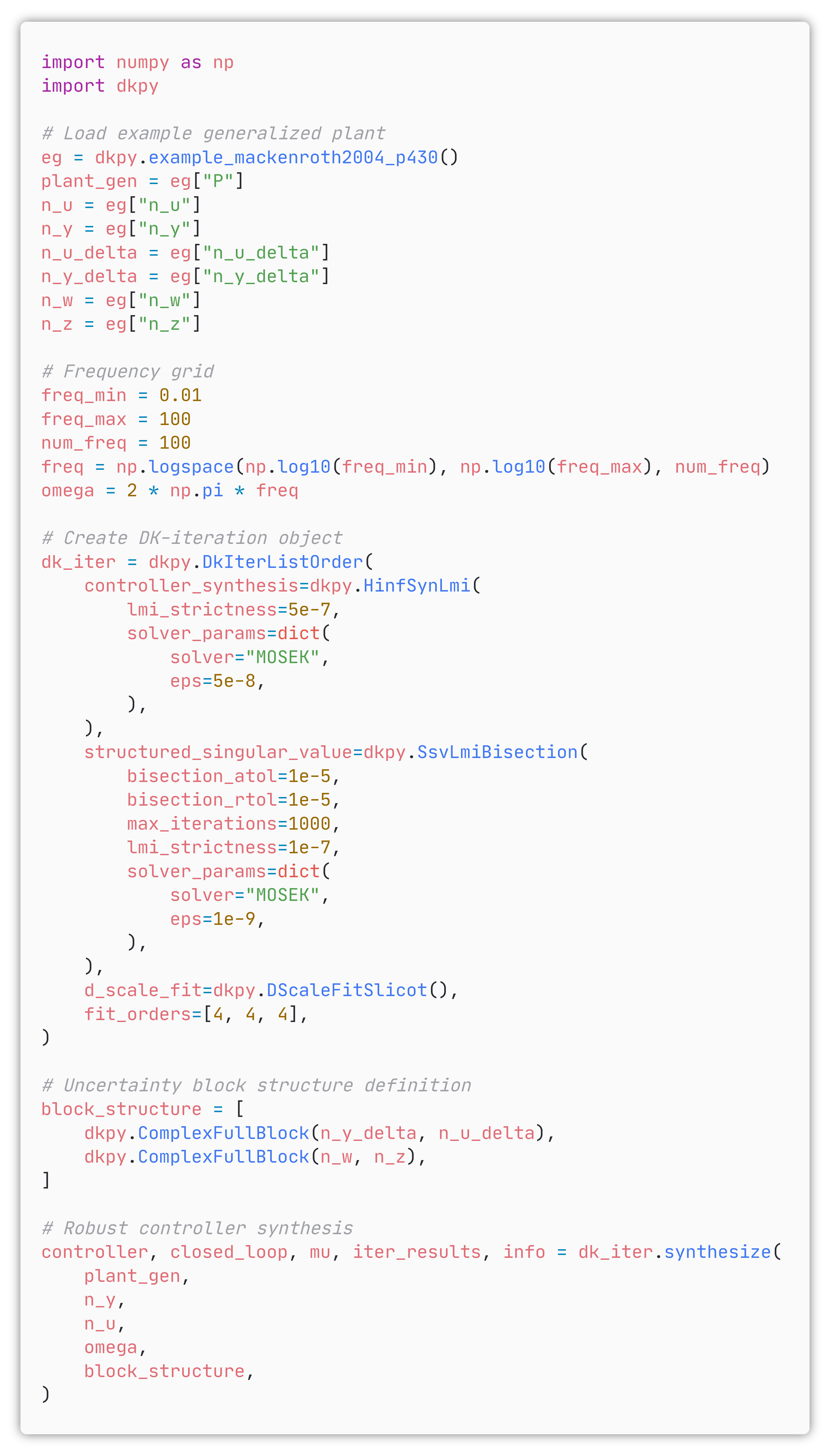}
    \caption{Controller synthesis code for robust control of the lateral
    dynamics of an aircraft using \texttt{dkpy}.}
    \vspace{-15pt}
    \label{fig:synthesis_example_code}
\end{figure}

\begin{figure}[t]
    \vspace{5pt}
    \centering
    \includegraphics[width=\linewidth]{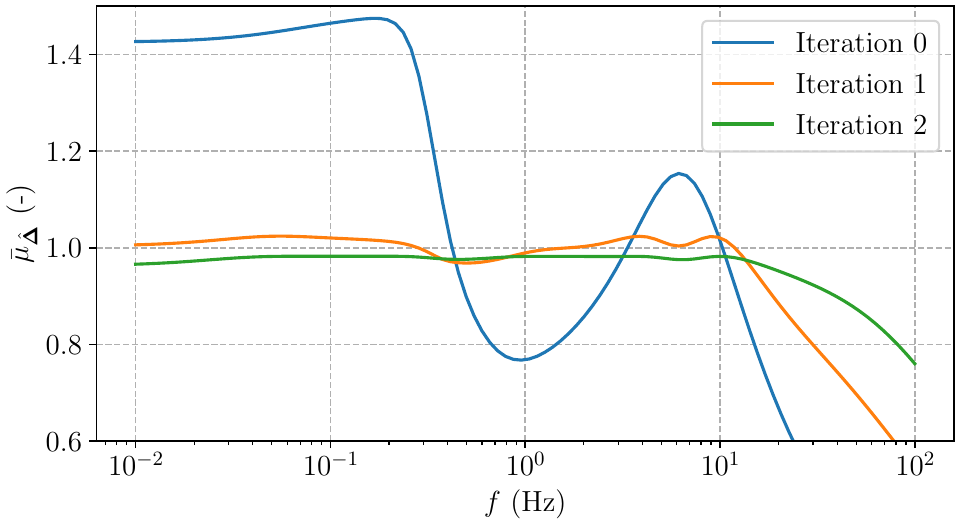}
    \caption{$DK$-iteration results of robust controller synthesis for lateral
    aircraft dynamics.}
    \vspace{-15pt}
    \label{fig:synthesis_example_dk_iteration_results}
\end{figure}

\begin{figure}[t!]
    \vspace{5pt}
    \centering
    \begin{subfigure}{\columnwidth}
        \centering
        \includegraphics[width=\linewidth]{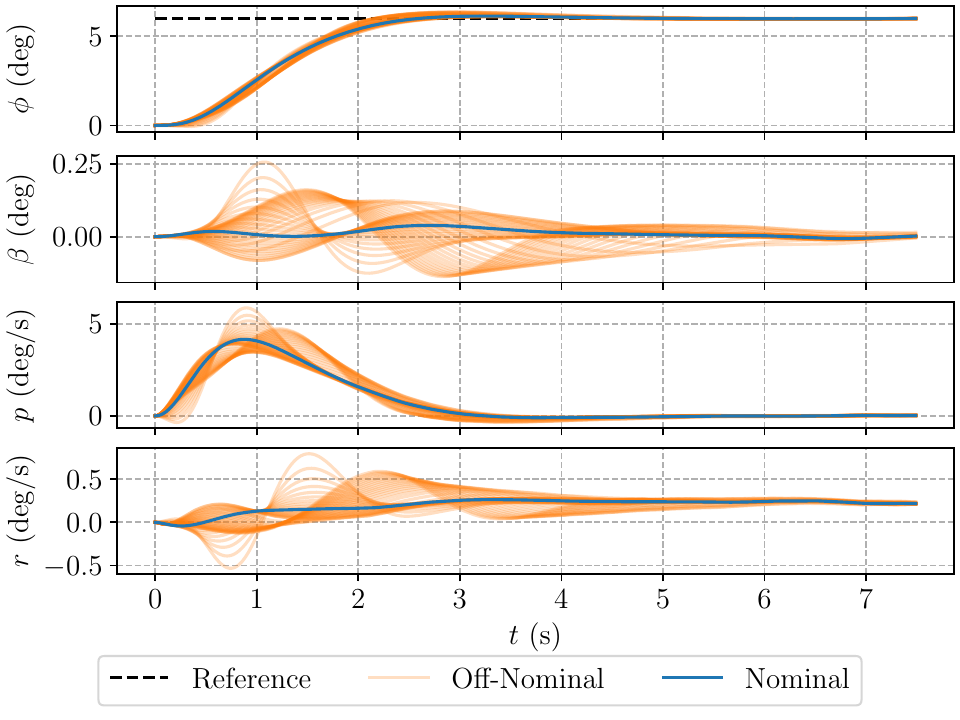}
        \caption{System states.}
        \label{fig:synthesis_example_state_response}
    \end{subfigure}
    \begin{subfigure}{\columnwidth}
        \centering
        \includegraphics[width=\linewidth]{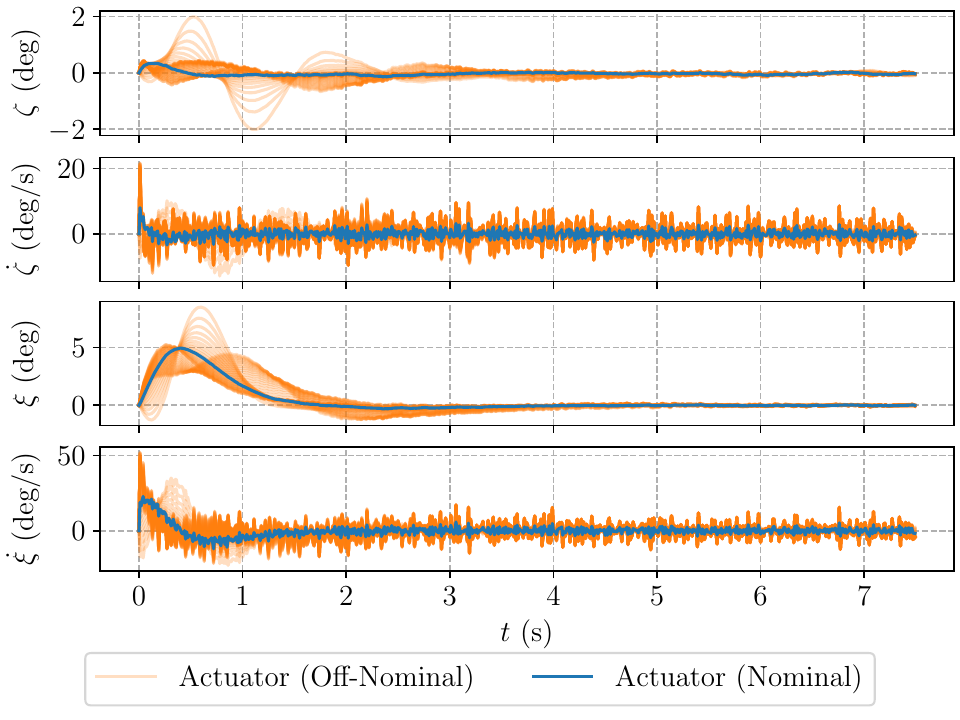}
        \caption{Control inputs.}
        \label{fig:synthesis_example_input_response}
    \end{subfigure}
    \caption{System response to a step in the reference roll angle $\phi_{r} =
    6 \, \si{(deg)}$ for the nominal and perturbed off-nominal models.}
    \label{fig:synthesis_example}
    \vspace{-15pt}
\end{figure}

The robust controller is synthesized with the \texttt{DkIterListOrder} class.
In this example, $3$ iterations are performed using $4$th order fits of
$\mbf{D}(s)$ for each iteration.
The robust performance condition is evaluated over a grid of $100$
logarithmically spaced frequencies between $0.01 \, (\si{Hz})$ and $100 \,
(\si{Hz})$.
The code required to perform the controller synthesis for the example is shown
in Figure \ref{fig:synthesis_example_code}.
The \texttt{HinfSynLmi} implementation is used for the $\mc{H}_{\infty}$
controller synthesis, \texttt{SsvLmiBisection} for the computation of the SSV
and scaling matrix frequency response, and \texttt{DScaleFitSlicot} for the
stable minimum phase fit of the scaling matrices.
The parameters for each of these methods are specified, which includes the
different numerical tolerances and optimization solvers used for each algorithm.
The example uses the \textit{MOSEK} optimization solver for the different
processes \cite{MOSEK}.

After completing the $DK$-iteration process, a controller $\mbf{K}(s)$ was
designed that realizes a maximum SSV of $0.982$ indicating that the robust
performance criterion is satisfied.
Figure~\ref{fig:synthesis_example_dk_iteration_results} shows the upper bound
of the SSV versus frequency across the iterations of $DK$-iteration.
It can be seen that the $DK$-iteration process minimizes the peak of the SSV
upper bound until it decreases below $1$ indicating that the controller design
satisfies the robust performance criterion.

Once the controller is designed, the closed-loop response may be simulated for 
a set of perturbed systems that are within the uncertainty set.
This is accomplished by forming the closed-loop system and selecting various
perturbations $\mbs{\Delta}(s)$ that satisfy $\norm{\mbs{\Delta}}_{\infty} \leq
1$.
To this end, $40$ different off-nominal systems are generated and the roll
angle reference step response is evaluated for $\phi_{r} = 6 \, (\si{deg})$.
The disturbances $\beta_{w}$ and $p_{w}$ are neglected and the sensor noise 
$n_{\phi}$, $n_{\beta}$, $n_{p}$, and $n_{r}$ are given by zero-mean white noise
with a standard deviation of $0.005 \, (\si{deg})$ and $0.005 \, (\si{deg/s})$.
Figure~\ref{fig:synthesis_example_state_response} shows the time response of
the states of the lateral aircraft dynamics subject to the roll angle reference
step for the nominal and perturbed off-nominal models.
The aircraft tracks the roll angle reference and that there is relatively
little variation between the nominal and off-nominal systems indicating good
robustness in the performance.
Figure~\ref{fig:synthesis_example_input_response} shows the time response of
the actuator inputs.
It can be seen that the rudder and aileron angles are relatively small with 
reasonable angular velocities indicating a reasonable control effort and that
the sensor noise has a noticeable but small effect on the control signals.
Moreover, the responses are consistent across the perturbed models, which
demonstrates the robustness of the controller design.

\section{Conclusion}
\label{sec:conclusion}

This paper presents \texttt{dkpy}, an open-source Python package for performing
robust controller analysis and synthesis with structured uncertainty using
$\mu$-analysis and $\mu$-synthesis.
%
% Moreover, \texttt{dkpy} provides tools for performing uncertainty
% characterization.
%
\texttt{dkpy} provides a modular interface for performing robust analysis and 
synthesis, which allows for greater flexibility for the user and an easier
method for exploring the various implementations and associated hyperparameters
to obtain the best results.

\printbibliography

\end{document}

%% file: figs/block_diagram/robust_stability_synthesis.tex
\newcommand{\internalblockspacing}[0]{0.3}
\newcommand{\externalblockspacing}[0]{1.5}
\newcommand{\horizontalspacing}[0]{0.5}

\begin{tikzpicture}[
    GeneralizedPlantNode/.style={
        rectangle,
        draw = black,
        thick,
        minimum height = 10mm,
        minimum width = 15mm
    },
    ControllerNode/.style={
        rectangle,
        draw = black,
        thick,
        minimum height = 8mm,
        minimum width = 15mm
    },
    UncertaintyNode/.style={
        rectangle,
        draw = black,
        thick,
        minimum height = 8mm,
        minimum width = 15mm
    },
]

% Generalized plant
\node[GeneralizedPlantNode]
    (GeneralizedPlant)
    {$\mbc{P}$};
\coordinate
    (GeneralizedPlantEastTop)
    at ($(GeneralizedPlant.east) + (0, \internalblockspacing)$);
\coordinate
    (GeneralizedPlantEastBottom)
    at ($(GeneralizedPlant.east) - (0, \internalblockspacing)$);
\coordinate
    (GeneralizedPlantWestTop)
    at ($(GeneralizedPlant.west) + (0, \internalblockspacing)$);
\coordinate
    (GeneralizedPlantWestBottom)
    at ($(GeneralizedPlant.west) - (0, \internalblockspacing)$);

% Uncertainty
\node[UncertaintyNode]
    (Uncertainty)
    at (0, \externalblockspacing)
    {$\mbs{\Delta}$};

% Controller
\node[ControllerNode]
    (Controller)
    at (0, -\externalblockspacing)
    {$\mbc{K}$};

\draw[->, thick]
    (GeneralizedPlantEastTop)
    -- ($(GeneralizedPlantEastTop) + (\horizontalspacing, 0)$)
    -- ($(Uncertainty.east) + (\horizontalspacing, 0)$)
    node [right, midway] {$\mbf{z}$}
    -- (Uncertainty.east);

\draw[->, thick]
    (Uncertainty.west)
    -- ($(Uncertainty.west) - (\horizontalspacing, 0)$)
    -- ($(GeneralizedPlantWestTop) - (\horizontalspacing, 0)$)
    node [left, midway] {$\mbf{w}$}
    -- (GeneralizedPlantWestTop);

\draw[->, thick]
    (GeneralizedPlantEastBottom)
    -- ($(GeneralizedPlantEastBottom) + (\horizontalspacing, 0)$)
    -- ($(Controller.east) + (\horizontalspacing, 0)$)
    node [right, midway] {$\mbf{y}$}
    -- (Controller.east);

\draw[->, thick]
    (Controller.west)
    -- ($(Controller.west) - (\horizontalspacing, 0)$)
    -- ($(GeneralizedPlantWestBottom) - (\horizontalspacing, 0)$)
    node [left, midway] {$\mbf{u}$}
    -- (GeneralizedPlantWestBottom);

\end{tikzpicture}

%% file: figs/block_diagram/robust_stability_analysis.tex
\newcommand{\internalblockspacing}[0]{0.2}
\newcommand{\externalblockspacing}[0]{1.5}
\newcommand{\horizontalspacing}[0]{0.5}

\begin{tikzpicture}[
    ClosedLoopNode/.style={
        rectangle,
        draw = black,
        thick,
        minimum height = 8mm,
        minimum width = 17.5mm
    },
    UncertaintyNode/.style={
        rectangle,
        draw = black,
        thick,
        minimum height = 8mm,
        minimum width = 17.5mm
    },
]

% Generalized plant
\node[ClosedLoopNode]
    (ClosedLoop)
    {$\mbc{F}_{\ell}(\mbc{P}, \mbc{K})$};

% Uncertainty
\node[UncertaintyNode]
    (Uncertainty)
    at (0, \externalblockspacing)
    {$\mbs{\Delta}$};

\draw[->, thick]
    (ClosedLoop.east)
    -- ($(ClosedLoop.east) + (\horizontalspacing, 0)$)
    -- ($(Uncertainty.east) + (\horizontalspacing, 0)$)
    node [right, midway] {$\mbf{w}$}
    -- (Uncertainty.east);

\draw[->, thick]
    (Uncertainty.west)
    -- ($(Uncertainty.west) - (\horizontalspacing, 0)$)
    -- ($(ClosedLoop.west) - (\horizontalspacing, 0)$)
    node [left, midway] {$\mbf{z}$}
    -- (ClosedLoop.west);

\end{tikzpicture}

%% file: figs/block_diagram/robust_performance_1.tex
\newcommand{\internalblockspacing}[0]{0.3}
\newcommand{\externalblockspacing}[0]{1.5}
\newcommand{\horizontalspacing}[0]{0.5}
\begin{tikzpicture}[
    GeneralizedPlantNode/.style={
        rectangle,
        draw = black,
        thick,
        minimum height = 10mm,
        minimum width = 15mm
    },
    ControllerNode/.style={
        rectangle,
        draw = black,
        thick,
        minimum height = 8mm,
        minimum width = 15mm
    },
    UncertaintyNode/.style={
        rectangle,
        draw = black,
        thick,
        minimum height = 8mm,
        minimum width = 15mm
    },
]

% Generalized plant
\node[GeneralizedPlantNode]
    (GeneralizedPlant)
    {$\mbc{P}$};
\coordinate
    (GeneralizedPlantEastTop)
    at ($(GeneralizedPlant.east) + (0, \internalblockspacing)$);
\coordinate
    (GeneralizedPlantEastBottom)
    at ($(GeneralizedPlant.east) - (0, \internalblockspacing)$);
\coordinate
    (GeneralizedPlantWestTop)
    at ($(GeneralizedPlant.west) + (0, \internalblockspacing)$);
\coordinate
    (GeneralizedPlantWestBottom)
    at ($(GeneralizedPlant.west) - (0, \internalblockspacing)$);

% Uncertainty
\node[UncertaintyNode]
    (Uncertainty)
    at (0, \externalblockspacing)
    {$\mbs{\Delta}$};

% Controller
\node[ControllerNode]
    (Controller)
    at (0, -\externalblockspacing)
    {$\mbc{K}$};

\draw[->, thick]
    (GeneralizedPlantEastTop)
    -- ($(GeneralizedPlantEastTop) + (\horizontalspacing, 0)$)
    -- ($(Uncertainty.east) + (\horizontalspacing, 0)$)
    node [right, midway] {$\mbf{z}_{1}$}
    -- (Uncertainty.east);

\draw[->, thick]
    (Uncertainty.west)
    -- ($(Uncertainty.west) - (\horizontalspacing, 0)$)
    -- ($(GeneralizedPlantWestTop) - (\horizontalspacing, 0)$)
    node [left, midway] {$\mbf{w}_{1}$}
    -- (GeneralizedPlantWestTop);

\draw[->, thick]
    (GeneralizedPlantEastBottom)
    -- ($(GeneralizedPlantEastBottom) + (\horizontalspacing, 0)$)
    -- ($(Controller.east) + (\horizontalspacing, 0)$)
    node [right, midway] {$\mbf{y}$}
    -- (Controller.east);

\draw[->, thick]
    (Controller.west)
    -- ($(Controller.west) - (\horizontalspacing, 0)$)
    -- ($(GeneralizedPlantWestBottom) - (\horizontalspacing, 0)$)
    node [left, midway] {$\mbf{u}$}
    -- (GeneralizedPlantWestBottom);

\draw[->, thick]
    ($(GeneralizedPlant.west) - (1.5*\horizontalspacing, 0)$)
    node [left, outer sep = -0.5mm] {$\mbf{w}_{2}$}
    -- (GeneralizedPlant.west);

\draw[->, thick]
    (GeneralizedPlant.east)
    -- ($(GeneralizedPlant.east) + (1.5*\horizontalspacing, 0)$)
    node [right, outer sep = -0.5mm] {$\mbf{z}_{2}$};

% Bounding box
% \draw [brown] (current bounding box.south west) rectangle (current bounding box.north east);
\end{tikzpicture}

%% file: figs/block_diagram/robust_performance_2.tex
\newcommand{\internalblockspacing}[0]{0.3}
\newcommand{\externalblockspacing}[0]{1.5}
\newcommand{\horizontalspacing}[0]{0.5}

\begin{tikzpicture}[
    GeneralizedPlantNode/.style={
        rectangle,
        draw = black,
        thick,
        minimum height = 10mm,
        minimum width = 15mm
    },
    ControllerNode/.style={
        rectangle,
        draw = black,
        thick,
        minimum height = 8mm,
        minimum width = 15mm
    },
    UncertaintyNode/.style={
        rectangle,
        draw = black,
        thick,
        minimum height = 8mm,
        minimum width = 15mm,
        inner sep=0mm,
    },
]

% Generalized plant
\node[GeneralizedPlantNode]
    (GeneralizedPlant)
    {$\mbc{P}$};
\coordinate
    (GeneralizedPlantEastTop)
    at ($(GeneralizedPlant.east) + (0, \internalblockspacing)$);
\coordinate
    (GeneralizedPlantEastBottom)
    at ($(GeneralizedPlant.east) - (0, \internalblockspacing)$);
\coordinate
    (GeneralizedPlantWestTop)
    at ($(GeneralizedPlant.west) + (0, \internalblockspacing)$);
\coordinate
    (GeneralizedPlantWestBottom)
    at ($(GeneralizedPlant.west) - (0, \internalblockspacing)$);

% Uncertainty
\node[UncertaintyNode]
    (Uncertainty)
    at (0, \externalblockspacing)
    {\scriptsize$\bbm \mbs{\Delta} & \mbf{0} \\ \mbf{0} & \mbs{\Delta}_{p} \ebm$};

% Controller
\node[ControllerNode]
    (Controller)
    at (0, -\externalblockspacing)
    {$\mbc{K}$};

\draw[->, thick]
    (GeneralizedPlantEastTop)
    -- ($(GeneralizedPlantEastTop) + (\horizontalspacing, 0)$)
    -- ($(Uncertainty.east) + (\horizontalspacing, 0)$)
    node [right, midway, outer sep = -1.25mm] {$\bbm \mbf{z}_{1} \\ \mbf{z}_{2} \ebm$}
    -- (Uncertainty.east);

\draw[->, thick]
    (Uncertainty.west)
    -- ($(Uncertainty.west) - (\horizontalspacing, 0)$)
    -- ($(GeneralizedPlantWestTop) - (\horizontalspacing, 0)$)
    node [left, midway, outer sep = -1.25mm] {$\bbm \mbf{w}_{1} \\ \mbf{w}_{2} \ebm$}
    -- (GeneralizedPlantWestTop);

\draw[->, thick]
    (GeneralizedPlantEastBottom)
    -- ($(GeneralizedPlantEastBottom) + (\horizontalspacing, 0)$)
    -- ($(Controller.east) + (\horizontalspacing, 0)$)
    node [right, midway] {$\mbf{y}$}
    -- (Controller.east);

\draw[->, thick]
    (Controller.west)
    -- ($(Controller.west) - (\horizontalspacing, 0)$)
    -- ($(GeneralizedPlantWestBottom) - (\horizontalspacing, 0)$)
    node [left, midway] {$\mbf{u}$}
    -- (GeneralizedPlantWestBottom);

% Bouding box
% \draw [brown] (current bounding box.south west) rectangle (current bounding box.north east);
\end{tikzpicture}

%% file: figs/block_diagram/generalized_plant.tex
\begin{tikzpicture}[
    SystemNode/.style={
        rectangle,
        draw = black,
        thick,
        minimum height = 8mm,
        minimum width = 15mm
    },
    ControllerNode/.style={
        rectangle,
        draw = black,
        thick,
        minimum height = 10mm,
        minimum width = 15mm
    },
    WeightNode/.style={
        rectangle,
        draw = black,
        thick,
        minimum height = 8mm,
        minimum width = 8mm
    },
    SummationNode/.style={
        circle,
        draw = black,
        thick,
        inner sep=0mm,
        minimum size = 2mm,
    },
]

% Airframe
\node[SystemNode]
    (Airframe)
    {$\mbc{G}_{\mathrm{AF}}$};
\coordinate
    (AirframeTopEast)
    at ($(Airframe.east) + (0, 0.2)$);
\coordinate
    (AirframeBottomEast)
    at ($(Airframe.east) - (0, 0.2)$);
\coordinate
    (AirframeTopWest)
    at ($(Airframe.west) + (0, 0.2)$);
\coordinate
    (AirframeBottomWest)
    at ($(Airframe.west) - (0, 0.2)$);

% Disturbance Weight
\node[WeightNode]
    (WeightDisturbance)
    % [above left=0.5 and 0.5 of Airframe]
    at ($(Airframe) + (-1.25, 1.25)$)
    {$\mbc{W}_{d}$};

% Input Derivative Performance Weight
\node[WeightNode]
    (WeightInputDerivative)
    % [left=0.5of WeightDisturbance]
    at ($(WeightDisturbance) + (-1.5, 0)$)
    {$\mbc{W}_{\dot{u}}$};

% Actuator
\node[SystemNode]
    (Actuator)
    at ($(WeightInputDerivative |- Airframe) + (-1.25, 0)$)
    {$\mbc{G}_{\mathrm{Act}}$};
\coordinate
    (ActuatorTopEast)
    at ($(Actuator.east) + (0, 0.2)$);
\coordinate
    (ActuatorBottomEast)
    at ($(Actuator.east) - (0, 0.2)$);

% Uncertainty summation junction
\node[SummationNode]
    (UncertaintySummation)
    at ($(Actuator) + (-1.5, 0)$)
    {};

% Delta block
\node[WeightNode]
    (DeltaBlock)
    at ($(UncertaintySummation) + (-0.75, 1.25)$)
    {$\mbs{\Delta}$};

% Uncertainty weight
\node[WeightNode]
    (UncertaintyWeight)
    at ($(DeltaBlock) + (-1.5, 0)$)
    {$\mbc{W}_{\Delta}$};

% Input performance weight
\node[WeightNode]
    (InputWeight)
    at ($(UncertaintyWeight) + (-1.5, 0)$)
    {$\mbc{W}_{u}$};
\coordinate
    (BelowInputWeight)
    at (InputWeight |- Airframe);
\coordinate
    (MidwayUncertaintyWeightInputWeight)
    at ($(InputWeight)!0.5!(UncertaintyWeight)$);
\coordinate
    (ToUncertaintyWeight)
    at (MidwayUncertaintyWeightInputWeight |- Airframe);

% Reference error difference 
\node[SummationNode]
    (ReferenceDifference)
    at ($(Airframe) + (2.0, 0.2)$)
    {};
\coordinate
    (Phi)
    at ($(AirframeTopEast)!0.5!(ReferenceDifference.west)$);

% Reference weight
\node[WeightNode]
    (ReferenceWeight)
    at ($(ReferenceDifference |- Airframe) + (0, 1.25)$)
    {$\mbc{W}_{r}$};

% Tracking Weight Node
\node[WeightNode]
    (TrackingWeight)
    at ($(ReferenceDifference |- Airframe) + (1.75, 0)$)
    {$\mbc{W}_{p}$};
\coordinate
    (TrackingWeightTopWest)
    at ($(TrackingWeight.west) + (0, 0.2)$);
\coordinate
    (TrackingWeightBottomWest)
    at ($(TrackingWeight.west) - (0, 0.2)$);
\coordinate
    (BetaPR)
    at ($(AirframeBottomEast)!0.5!(TrackingWeightBottomWest)$);

% Noise weight
\node[WeightNode]
    (NoiseWeight)
    at ($(TrackingWeight) + (0, -1.75)$)
    {$\mbc{W}_{n}$};
\coordinate
    (NoiseWeightTopWest)
    at ($(NoiseWeight.west) + (0, 0.2)$);
\coordinate
    (NoiseWeightBottomWest)
    at ($(NoiseWeight.west) - (0, 0.2)$);

% Noise summation
\node[SummationNode]
    (NoiseSummationPhi)
    at (Phi |- NoiseWeightBottomWest)
    {};
\node[SummationNode]
    (NoiseSummationRest)
    at (BetaPR |- NoiseWeightTopWest)
    {};

% Controller
\coordinate
    (CenterActuatorAirframe)
    at ($(Actuator)!0.5!(Airframe)$);
\node[ControllerNode]
    (Controller)
    at (CenterActuatorAirframe |- NoiseSummationRest)
    {$\mbc{K}$};
\coordinate
    (ControllerTopEast)
    at ($(Controller.east) + (0, 0.4)$);
\coordinate
    (ControllerBottomEast)
    at ($(Controller.east) - (0, 0.4)$);

\draw[->, thick]
    (ActuatorBottomEast)
    -- (AirframeBottomWest)
    node[below, midway] {$(\zeta, \xi)$};

\draw[->, thick]
    (ActuatorTopEast)
     -| (WeightInputDerivative.south)
     node [right, pos=0.75] {$(\dot{\zeta}, \dot{\xi})$};

\draw[->, thick]
    (WeightInputDerivative.north)
    -- ($(WeightInputDerivative.north) + (0, 0.5)$)
    node [right] {$\mbf{z}_{\dot{u}}$};

\draw[->, thick]
    (WeightDisturbance.south)
     |- (AirframeTopWest);

\draw[->, thick]
    ($(WeightDisturbance.north) + (0, 0.5)$)
    node [right] {$(\beta_{w}, p_{w})$}
    -- (WeightDisturbance.north);

\draw[->, thick]
    (UncertaintySummation.east)
    -- (Actuator.west);

\draw[->, thick]
    (UncertaintyWeight.east)
    -- (DeltaBlock.west)
    node [above, midway] {$\mbf{y}_{{\Delta}}$};

\draw[->, thick]
    (DeltaBlock.east)
    -| (UncertaintySummation.north)
    node [above, pos=0.5] {$\mbf{u}_{{\Delta}}$};

\draw[->, thick]
    (InputWeight.north)
    -- ($(InputWeight.north) + (0, 0.5)$)
    node [right] {$\mbf{z}_{u}$};

\draw[->, thick]
    (ReferenceWeight.south)
    -- (ReferenceDifference.north);

\draw[->, thick]
    ($(ReferenceWeight.north) + (0, 0.5)$)
    node[right] {$\phi_{r}$}
    -- (ReferenceWeight.north);

\draw[->, thick]
    ($(Airframe.east) + (0, 0.2)$)
    -- (ReferenceDifference.west)
    node[below, pos=0.9] {$-$}
    node[above, pos=0.5] {$\phi$};

\draw[->, thick]
    (ReferenceDifference.east)
    -- (TrackingWeightTopWest);

\draw[->, thick]
    (AirframeBottomEast)
    -- (TrackingWeightBottomWest)
    node [below, pos=0.75] {$(\beta, p, r)$};

\draw[->, thick]
    (TrackingWeight.east)
    -- ($(TrackingWeight.east) + (0.5, 0)$)
    node [right] {$\mbf{z}_{p}$};

\draw[->, thick]
    (Phi)
    -- (NoiseSummationPhi.north);

\draw[->, thick]
    (BetaPR)
    -- (NoiseSummationRest.north);

\draw[->, thick]
    (NoiseWeightBottomWest)
    -- (NoiseSummationPhi);

\draw[->, thick]
    (NoiseWeightTopWest)
    -- (NoiseSummationRest);

\draw[->, thick]
    ($(NoiseWeight.east) + (0.5, 0)$)
    node [right] {$\mbf{n}$}
    -- (NoiseWeight.east);

\draw[->, thick]
    (NoiseSummationPhi.west)
    -- (ControllerBottomEast);
\draw[->, thick]
    (NoiseSummationRest.west)
    -- (Controller.east);
\draw[->, thick]
    ($(ControllerTopEast) + (0.5, 0)$)
    node [right] {$\phi_{r}$}
    -- (ControllerTopEast);

\draw[->, thick]
    (Controller.west)
    -| (InputWeight.south)
    node [below, pos=0.25] {$(\zeta_{c}, \xi_{c})$};

\draw[->, thick]
    (BelowInputWeight)
    -- (UncertaintySummation.west);

\draw[->, thick]
    (ToUncertaintyWeight)
    |- (UncertaintyWeight.west);

\end{tikzpicture}